\documentclass[aps,prl,floatfix,twocolumn]{revtex4-2}
\usepackage[dvips]{graphicx}

\usepackage{longtable}
\usepackage{dcolumn}
\usepackage[dvips]{graphicx}
\usepackage{bm}
\usepackage{bbm}
\usepackage{slashed}
\usepackage{times}
\usepackage{nicefrac}
\usepackage{amsmath}
\usepackage{amsfonts}
\usepackage{amssymb}
\usepackage{amsthm}

\newcolumntype{.}{D{x}{}{-1}}
\newcolumntype{w}[1]{D{.}{.}{#1}}

\usepackage{color}

\newcommand{\vare}{\varepsilon}

\newcommand{\lbr}{\langle}
\newcommand{\rbr}{\rangle}

\newcommand{\Za}{Z\alpha}

\allowdisplaybreaks

\begin{document}

\title{QED theory of the nuclear recoil with  finite size}

\author{Krzysztof Pachucki}
\affiliation{Faculty of Physics, University of Warsaw,
             Pasteura 5, 02-093 Warsaw, Poland}

\author{Vladimir A. Yerokhin}
\affiliation{Peter the Great St.~Petersburg Polytechnic University,
Polytekhnicheskaya 29, 195251 St.~Petersburg, Russia}

\begin{abstract}

We investigate the modification of the transverse electromagnetic interaction
between two point-like particles when one particle acquires a finite size. It is
shown that the correct treatment of such interaction
cannot be accomplished within the Breit approximation but should be
addressed within the QED. The complete QED formula is derived
for the finite-size nuclear recoil, exact in the coupling strength parameter $\Za$.
Numerical calculations are carried out for a wide range of $Z$ and
verified against the $(\Za)^5$ contribution. The comparison
with the $\Za$ expansion identifies the contribution of order $(\Za)^6$,
which is linear in the nuclear radius and numerically dominates over the
lower-order $(\Za)^5$ term.

\end{abstract}

\maketitle

{\sl Introduction.---}The
relativistic spin-$\nicefrac{1}{2}$ particle in the Coulomb field of the infinitely heavy nucleus is described by the Dirac equation.
In contrast to the nonrelativistic case, the finite nuclear mass effects,
often called the nuclear recoil, cannot be incorporated into the Dirac equation but
should be addressed within QED theory. The QED calculations of the nuclear recoil started with
pioneering works of Salpeter~\cite{salpeter:52} in 1952. In late 1980s it was proven
\cite{shabaev:85,shabaev:88,pachucki:95,yelkhovsky:96,shabaev:98:rectheo} that the linear in $m/M$
nuclear recoil can be described by a closed-form formula
valid to all orders in the electron-nucleus coupling strength $\Za$
(where $m$ is the electron mass, $M$ is the nuclear mass, $Z$ is the nuclear charge number and $\alpha$ is the fine-structure
constant). The  numerical calculations were performed in Refs.~\cite{artemyev:95:pra,artemyev:95:jpb}.
Later, this formula has been generalized for the external homogenous magnetic field, which stimulated
extensive research on the bound electron g-factor \cite{koehler:16,sailer:22}.
All these studies considered the nuclear recoil effect only for the point-like nucleus.

Let us now consider a finite-size nucleus. The modification of the electrostatic potential by the finite
size is straightforward and can be immediately incorporated in the Dirac
equation. By contrast, the corresponding modification of the nuclear recoil turns out
to be highly nontrivial and has not yet been properly performed in the literature. Within the Breit approximation,
the finite-size nuclear recoil correction was derived by Borie and Rinker \cite{borie:82} and later
rederived in Ref.~\cite{aleksandrov:15}. In this Letter we show that these derivations were incomplete
and obtain the exact formula for the  Breit interaction for a finite nucleus.
It is remarkable that the Breit-approximation
formula, even the correct one, should not be used for comparison with experiment since it contains
\cite{shabaev:98:recground} a spurious contribution
$\sim \! r_C\,(\Za)^5m/M$
linear in the nuclear charge radius
that overshadows the main contribution
$\sim \! r_C^2\,(\Za)^4m/M$. The correct handling of the finite-size nuclear recoil is possible only within the QED.

In this Letter we obtain the complete formula for the finite-size nuclear recoil correction and
perform numerical calculations for the whole range of $Z$. Within an alternative approach, we derive the
contribution of order $(\Za)^5m/M$. The comparison of the all-order (in $\Za$)
results with the $\Za$-expansion calculations gives us access to the contribution of order $(\Za)^6m/M$,
which is linear in the nuclear charge radius and numerically dominates over the previous-order
contribution.

{\sl Expansion in the small nuclear charge.---}
Let us denote by $E_{\rm fns}$ the shift in the binding energy of a hydrogenic system due to the finite nuclear size (fns).
For a light atom we can perform the expansion of $E_{\rm fns}$ in the small nuclear charge
\begin{align}
 E_{\rm fns} =  E^{(4)}_{\rm fns} +  E^{(5)}_{\rm fns}+  E^{(6)}_{\rm fns} + \ldots
 \end{align}
where the superscript indicates the order in $\Za$.
The leading-order nuclear contribution is of order $(\Za)^4$ and given by a simple formula,
\begin{equation}
  E^{(4)}_{\rm fns} =  \frac{2\,\pi}{3}\,\Za\,\phi^2(0)\,r_C^2\,, \label{02}
\end{equation}
where $\phi(0)$ is the nonrelativistic wave function of the electron at the position of nucleus,
$r_C$ is the root-mean-square charge radius of the nucleus,
$ r_C^2 = \int d^3 r\, r^2\,\rho(\vec r)$,
and $\rho(\vec r)$ is the nuclear charge distribution. Eq.~(\ref{02}) includes the exact dependence on
the finite nuclear mass $M$
through $\phi^2(0) = m_r^3\,(\Za)^3/(\pi n^3)$, where $m_r = mM/(m+M)$.

{\sl $(\Za)^5$ finite nuclear size.---}
The description of fns effects for an arbitrary mass ratio at the order $(\Za)^5$ is much more complicated.
We here thus briefly discuss the approximations and assumptions needed to derive
this correction. Let us start from the general expression for the
nuclear-structure contribution of order $(\Za)^5$,
\begin{align}\label{eq3}
 E^{(5)}_{\rm nucl} =& -\frac{(Z\,e^2)^2}{2}\, \phi^2(0)\,\int\frac{d^4q}{(2\,\pi)^4\,i}\,
\frac{1}{q^4}
\nonumber \\ & \times
\big[T^{\mu\sigma}(I,M)-t^{\mu\sigma}(I,M)\big]\,t_{\mu\sigma}(\nicefrac12,m)\,,
\end{align}
where $T^{\mu\sigma}(I,M)$ and $t^{\mu\sigma}(\nicefrac12,m)$ are the forward virtual Compton scattering amplitudes off
the nucleus (with the spin $I$ and mass $M$), and the electron (with the spin $\nicefrac12$ and mass $m$), respectively.
Furthermore,
$t^{\mu\sigma}(I,M)$ is the point-nucleus limit of $T^{\mu\sigma}(I,M)$.
The subtraction of the point-nucleus limit in above equation is necessary because it is already included into the $(\Za)^5$
nuclear recoil correction \cite{salpeter:52, pachucki:23}.
For the electron, the scattering amplitude is very simple and given by
\begin{align}
t^{\mu\sigma}(\nicefrac12,m) =&\
    {\rm Tr}\biggl[\gamma^\mu\frac{1}{m\!\not\!t+\not\!q-m}\gamma^\sigma\,\frac{\gamma^0+I}{4}\biggr] + (q\rightarrow-q)\,, \label{04}
\end{align}
with $\nu = q^0$ and $t=(1,0,0,0)$. By contrast, for the nuclear scattering amplitude $T^{\mu\nu}$ we usually do
not have much information.
Nevertheless, the gauge invariance requires that $q_\mu\,T^{\mu\sigma}=0$ and therefore
$T^{\mu\sigma}$ can be expressed in terms of only  two Lorentz invariant functions $T_1$ and $T_2$,
\begin{align}
T^{\mu\sigma} =&
-\biggl(g^{\mu\sigma}-\frac{q^\mu\,q^\sigma}{q^2}\biggr)\,\frac{T_1}{M}
\nonumber \\ &\
+\biggl(t^\mu-\frac{\nu}{q^2}\,q^\mu\biggr)\,
\biggl(t^\sigma-\frac{\nu}{q^2}\,q^\sigma\biggr)\,\frac{T_2}{M}\,.
\end{align}
Using this parametrization, we evaluate Eq.~(\ref{eq3}) as
\begin{align}
E^{(5)}_{\rm nucl} &\ =
-2\,(Z\,e^2)^2\,\phi^2(0)\,\frac{m}{M}\,\int\frac{d^4q}{(2\,\pi)^4\,i}\nonumber \\ &\hspace*{-5ex}\times
\frac{[T_2-t_2(I,M)](q^2-\nu^2)-[T_1-t_1(I,M)]\,(q^2+2\,\nu^2)}
{q^4\,(q^4-4\,m^2\nu^2)}\,,
\end{align}
where $t_1$ and $t_2$ are the point-nucleus limits of $T_1$ and $T_2$, respectively.

Let us now split the nuclear contribution into the fns and polarizability parts,
$E^{(5)}_{\rm nucl} = E^{(5)}_\mathrm{fns} + E^{(5)}_\mathrm{pol}$ .
The separation is not unique and was carried out in different ways in the literature. We
here separate the fns part by assuming that nucleus is described only by the elastic formfactors; this definition
is often referred to as the Born contribution.
For the spin-zero nuclei, there is only the charge formfactor $\rho(-q^2)$.
For an arbitrary spin $I$, there are in addition the magnetic, quadrupole and possibly other formfactors.
However,  to the zeroth and the first order in $m/M$ only the charge formfactor contributes.
Under this assumption, the fns contribution becomes
\begin{align}
E^{(5)}_{\rm fns} = &\
-2\,(Z\,e^2)^2\,\phi^2(0)\,\frac{m}{M}\,\int\frac{d^4q}{(2\,\pi)^4\,i}\,\big[\rho^2(-q^2)-1\big]\nonumber \\ &\times
\frac{t_2(I,M)(q^2-\nu^2)-t_1(I,M)\,(q^2+2\,\nu^2)}
{q^4\,(q^4-4\,m^2\nu^2)}\,.
\end{align}
We now claim that the nonrecoil and the leading recoil corrections do not depend on the nuclear spin,
which allows us to set  $I=1/2$ and obtain $t_1, t_2$ from Eq. (\ref{04}).
Next we perform the angular integration in the Euclidean momentum space,
\begin{align}
E^{(5)}_{\rm fns} =&\  -(\Za)^2\,\phi^2(0)\,m \int_0^\infty \frac{dp}{p} \, T(p^2)\,,
\end{align}
and expand $T(p^2)$ in large $M$ as
\begin{align}
T(p^2) =&\  T^{(0)}(p^2) + \frac{T^{(1)}(p^2)}{M} + O\Bigl(\frac{1}{M}\Bigr)^2\,.
\end{align}
The leading term
$T^{(0)} = (16/p^3)\,\big[\rho^2(p^2) -1-2\,p^2\,\rho'(0)\big]$
corresponds to the non-recoil limit. Performing the momentum integration as
\begin{align}
\int_0^\infty\frac{dp}{p}\,T^{(0)} =&\  r_F^3\,\frac{\pi}{3}\,,
\end{align}
where $r_F^3 =  \int d^3r_1\int d^3r_2\,\rho(r_1)\,\rho(r_2)\,|\vec r_1-\vec r_2|^3$,
we reproduce the well-known Friar correction
\cite{friar:79:ap},
\begin{align}
  E^{(5)}_{\rm fns}(M = \infty) = -\frac{\pi}{3}\,\phi^2(0)\,(\Za)^2\,m\,r_F^3\,.\label{13}
\end{align}
The leading recoil term  in expansion of $T$ in the mass ratio is
\begin{align}
T^{(1)} =&\ \frac{8}{p^2}\,\Big[\sqrt{1 + a^2} - \big(1 + \sqrt{1 + a^2}\big)^{-2}\Big]\big[1 - \rho^2(p^2)\big]
 \nonumber \\ &
  + 16\,a\,\rho'(0)\,,
\end{align}
where $a=2\,m/p$. The momentum integral is represented in the coordinates space as
\begin{align}
\int_0^\infty\frac{dp}{p}\,T^{(1)} =&\ \Big[\frac{7}{6} - 2\,\gamma - 2\,\ln(m\,r_L)\Big]\,r_C^2\,, \label{15}
\end{align}
with the effective radius $r_L$ defined by
\begin{align}
\int d^3r_1\int d^3r_2\,\rho(\vec r_1)\,\rho(\vec r_2)\,|\vec r_1-\vec r_2|^2\, & \ln(m\,|\vec r_1-\vec r_2|) \nonumber \\
\equiv&\  2\,r_C^2\,\ln(m\,r_L)\,. \label{16}
\end{align}
Finally, the fns recoil correction of order $(\Za)^5$ is
\begin{align}\label{eq:17}
  E^{(5)}_{\rm recfns} = &\   -\frac{m}{M}\,\phi^2(0)\,(Z\,\alpha)^2\,\Big[ \frac{7}{6} - 2\,\gamma - 2\,\ln(m\,r_L)\Big]\,r_C^2 \,,
\end{align}
where we omitted the reduced-mass correction in Eq.~(\ref{13})
since it is two orders of magnitude smaller for normal "electronic" atoms.
The effective radius $r_L$ for the exponential model amounts to $1.74\,r_C$ and
should not significantly differ for other nuclear charge distributions.
In comparison to the leading fns effect given by Eq.~(\ref{02}), $E^{(5)}_{\rm recfns}$ is
decreased by a factor of $\Za\,m/M$ but enhanced by $\ln(m\,r_L)$, which is  $\approx -5.6$ for hydrogen.\\

{\sl $(\Za)^5$ effects beyond the finite nuclear size.---}
It is well known that the treatment of a nucleus as a finite-size particle omits numerous nuclear-structure
effects, often termed as the nuclear polarizability contribution.
Subtracting from $T_1$ and $T_2$ the fns parts, one writes the nuclear polarizability correction as
\begin{align}
E^{(5)}_\mathrm{pol} =&
-2\,(Z\,e^2)^2\,\phi^2(0)\,\frac{m}{M}\,\int\frac{d^4q}{(2\,\pi)^4\,i}\nonumber \\ & \times
\frac{T_{2}\,(q^2-\nu^2)-T_{1}\,(q^2+2\,\nu^2)}
{q^4\,(q^4-4\,m^2\nu^2)}\,. \label{67}
\end{align}
An approach used in the literature is
to employ dispersion relations in the variable $\nu$ to express $T_1(\nu,-q^2)$ and $T_2(\nu,-q^2)$  in terms of structure functions
that in principle can be measured  in the electron-nucleus scattering.
In the case of $T_1$, a subtracted dispersion relation is needed,
giving rise to the subtraction function $T_{1}(0,-q^2)$, which can not be measured directly
but needs to be calculated from the nuclear theory,
with a condition that its small-$q$ behavior is governed by the  magnetic dipole polarizability $T_1 = \alpha/M\,q^2\,\beta_M + O(q^4)$.
The structure functions are known experimentally only for the proton, deuteron, and helion,
and only for a part of the kinematic space. Generally, usage of dispersion relations for nuclei
heavier than proton requires significant input from the nuclear theory. Such calculations
were recently performed for the deuteron in Refs.~\cite{acharya:21,lensky:22}.

An alternative approach is to
calculate  the total nuclear structure correction by considering the nucleus as
a system of individual interacting nucleons and do not introduce the fns effect at all. Such calculations are
nowadays feasible for light nuclei.
Specifically, the nuclear contribution of order $(\Za)^5$ for a light composite nucleus is written as
\cite{pachucki:18}
\begin{align}
  E^{(5)}_{\rm nucl} =&\ E^{(5)}_{\rm nucl1} + E^{(5)}_{\rm nucl2} + E^{(5)}_{\rm pol}\,, \label{38}\\
  E^{(5)}_{\rm nucl1} =&\ -\frac{\pi}{3}\,m\alpha^2\phi^2(0)
  \Big[Z\,R_{pF}^3 + (A-Z)\,R_{nF}^3 \Bigr]\,,\\
  E^{(5)}_{\rm nucl2} =&\ -\frac{\pi}{3}\,m\alpha^2\phi^2(0)
  \sum_{i,j=1}^Z\langle\phi_N||\vec R_i-\vec R_j|^3|\phi_N\rangle\,. \label{40}
\end{align}
Here $E^{(5)}_{\rm nucl1}$ comes from the two-photon exchange with the same nucleon, $E^{(5)}_{\rm nucl2}$ is due to the two-photon exchange
with different nucleons, and $E^{(5)}_{\rm pol}$ is the nuclear polarizability correction originating from the low-energy two-photon exchange.
The parameters $R_{pF}$ and $R_{nF}$ are the effective proton and neutron radii, correspondingly. They represent
the complete two-photon exchange (with subtracted point-proton contribution) and thus include the recoil with
individual nucleons. We extract them from the calculation of Tomalak~\cite{tomalak:19}, with the result
$R_{pF} = 1.947\,(75)\,{\rm fm}$ and $R_{nF} = 1.43\,(16)\;{\rm fm}$.

Unfortunately, it is not feasible at present to extend this approach to nuclei consisting of many nucleons or to
effects of higher orders in $\Za$. For complex nuclei, the only currently available way is to assume the charge
form factor model and separately account for the nuclear polarizability effects as was done in Refs.~\cite{plunien:95,nefiodov:96}.
We thus return to the description of nucleus through the elastic charge formfactor, but keep in mind
the limitations of this very simplified picture.

{\sl Photon propagator in the modified Coulomb gauge.---}
In order to obtain a formula for the relativistic recoil correction that is valid for an arbitrary $Z$,
we shall construct the photon propagator  with one finite-size vertex in the Coulomb gauge.
First we consider the Feynman gauge. In this case
the photon propagator with the charge formfactor is given by
\begin{align}
G_F^{\mu\sigma}(k) = -\frac{g^{\mu\sigma}}{k^2}\,\rho(-k^2)\,,
\end{align}
where we assumed that the formfactor can be analytically continued into the complex plane
with possible poles and branch cuts on the negative real axis $-k^2 < 0$.
In the Coulomb gauge we require that the scalar part of the propagator
coincides with the Coulomb potential of an extended nucleus, namely
$G_C^{00} =\rho(\vec k^2)/\vec k^2$.
Then the transverse part of the propagator has to be of the form
\begin{align}
G_C^{ij}(k) =&\ \frac{\rho(-k^2)}{k^2}\,\biggl(\delta^{ij}-\frac{k^i\,k^j}{(k^0)^2}\biggr) - \frac{k^i\,k^j}{(k^0)^2}\,\frac{\rho(\vec k^2)}{\vec k^2}\,.
\label{eq:23}
\end{align}
The above formula is justified by the equivalence of $G_F$ and $G_C$ that follows from the gauge transformation
\begin{align}
G_F^{\mu\sigma} =&\ G_C^{\mu\sigma} + k^\mu f^\sigma + f^\mu\,k^\sigma\,,
\end{align}
with $f^0 = -k^0f$, $f^i = k^if$, and
\begin{align}
f =&\ \frac{1}{2\,(k^0)^2}\bigg[\frac{\rho(\vec k^2)}{\vec k^2} + \frac{\rho(-k^2)}{k^2}\bigg]\,.
\end{align}

The coordinate-space representation of the transverse part of the propagator is obtained as
\begin{align}
G_{C}^{ij}(\omega,\vec{r}) = \delta^{ij}\,{\cal D}(\omega,r) + \frac{\nabla^i\nabla^j}{\omega^2} \, \Big[ {\cal D}(\omega,r)
 - {\cal D}(0,r)\Big]\,,
\end{align}
where $\omega \equiv k^0$ and
\begin{align}
{\cal D}(\omega,r) = \int \frac{d^3k}{(2\pi)^3}\, e^{i\vec{k}\cdot\vec{r}}\,\frac{\rho({\vec k}^2-\omega^2)}{\omega^2-{\vec k}^2}\,.
\end{align}
The Breit-approximation formula for the transverse electron-nucleus interaction is obtained
by taking the limit $\omega \to 0$, with the result
\begin{align}
G_C^{ij}(0,\vec{r})  =
\frac{1}{2}\,\biggl(\delta^{ij} -\frac{r^ir^j}{r}\,\frac{d}{dr}\biggr)\,{\cal D}(0,r)\,.
\end{align}
It coincides with the result obtained previously in Ref.~\cite{veitia:04} but disagrees with the later work~\cite{aleksandrov:15}.

{\sl Finite-size nuclear recoil for an arbitrary nuclear charge.---}
In order to obtain the finite-size nuclear recoil correction
we use the formula originally derived for the point nucleus to all orders in $Z\alpha$
\cite{shabaev:85,shabaev:88,pachucki:95,yelkhovsky:96,shabaev:98:rectheo}
and replace the point-nucleus photon propagator in the Coulomb gauge by the finite-nucleus photon propagator.
This procedure can be justified by considering the electron-nucleus scattering amplitude. Every photon exchange
is described by the propagator $-g^{\mu\nu}/k^2$ and a formfactor vertex on the nucleus line.
Performing the nonrelativistic limit for the nucleus, we arrive at the scattering amplitude of point-like nonrelativistic particles
that interact by means of the modified photon propagator. 
The nuclear recoil correction was derived assuming the nonrelativistic Hamiltonian for a point nucleus,
thus for the finite-size nucleus we obtain
\begin{align}\label{eq:rec}
E_{\rm rec} = \frac{m^2}{M}\frac{i}{2\pi}\, \int_{-\infty}^{\infty}
  &\
d\omega\,
 \lbr a | \big[ p^j - D^j(\omega)\big] \,
  \nonumber \\ & \times
 G(\omega + \vare_a) \,
 \big[ p^j - D^j(\omega)\big] | a \rbr\,,
\end{align}
where  $G(E) = [E-H_D(1-i\epsilon)]^{-1}$ is the Dirac-Coulomb Green function,
$D^j(\omega) = -4\pi Z\alpha \, \alpha^i \, G_{C}^{ij}(\omega,\vec{r})$,
and $\alpha^i$ are the Dirac matrices.

In order to proceed further we need to specify explicitly the model of the
nuclear charge distribution. We will
use the exponential model, whose kernel in the momentum space is
$\rho({\vec k}^2) = \lambda^4/(\lambda^2+{\vec k}^2)^2$,
where $\lambda = 2\sqrt{3}/r_C$.
Since the recoil correction (\ref{eq:rec}) is calculated after the Wick rotation $\omega \to i\omega$
(see Ref.~\cite{artemyev:95:pra}), for performing
calculations for the $1s$ reference state we need the photon propagator for imaginary energies only. We obtain
for $\omega = i\omega_+$ and $\omega_+\geq0$,
\begin{align}
{\cal D}(i\omega_+,r) = -\frac1{4\pi}\, \left[ \frac{e^{-\omega_+\, r}}{r} - \frac{e^{-\overline{\omega}_+ \,\, r}}{r}
 - \frac{\lambda^2}{2}\frac{e^{-\overline{\omega}_+  \,\,r}}{\overline{\omega}_+ }
 \right]\,,
\end{align}
where $\overline{\omega}_+ = \big(\omega_+^2+\lambda^2\big)^{1/2}$ and
${\cal D}(-i\omega_+,r) = {\cal D}(i\omega_+,r)$.

\begin{figure}
\centerline{
\resizebox{\columnwidth}{!}{%
  \includegraphics{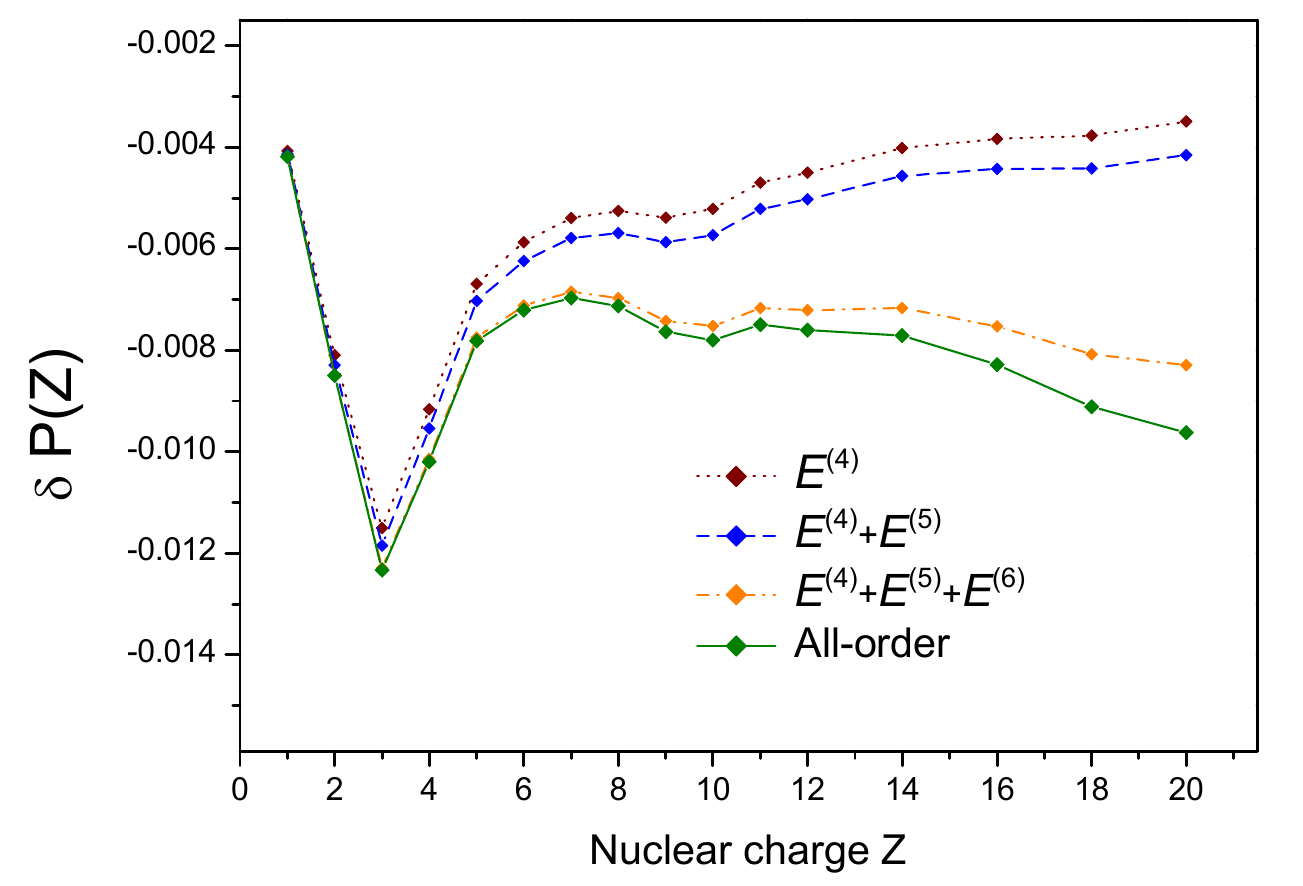}
}}
 \caption{
 Finite-size nuclear recoil correction for the $1s$ state of H-like ions,
 in terms of function $\delta P = E_{\rm recfns}/[(m^2/M) (\Za)^5/\pi]$.
 \label{fig:1}}
\end{figure}

We performed numerical calculations of the finite-size nuclear recoil correction
to all orders in $\Za$ by evaluating Eq.~(\ref{eq:rec}) for the extended and the point
nuclear models and taking the difference. Results of our numerical calculations are shown in Fig.~\ref{fig:1}, in
comparison with contributions of the $\Za$-expansion corrections. The plotted function depends both on $Z$
and $r_C$, leading to a non-smooth behaviour of the plots in Fig.~\ref{fig:1}. We observe that the
sum $E^{(4)}_{\rm recfns} + E^{(5)}_{\rm recfns}$ differs noticeably from the all-order results
already for moderate values of $Z$. By varying separately $Z$ and $r_C$ in our numerical calculations,
we determined that the reason is the contribution of the next order in $(\Za)$, which depends
-- very unusually -- linearly on $r_C$. We thus deduce the contribution of order $(\Za)^6$
of the form
\begin{align}\label{E6}
E^{(6)}_{\rm recfns} = -\frac{m^3}{M}\, a^{(6)} \,(\Za)^6\,r_C\,,
\end{align}
where the numerical value of the coefficient $a^{(6)} \approx 1.0$.
This approximate equation is obtained for the exponential nuclear model;
for other models we might expect a different effective radius instead of $r_C$, but
the linear dependence shall remain. Fig.~\ref{fig:1} demonstrates that
the inclusion of the $(\Za)^6$ contribution significantly improves agreement between the all-order and $\Za$-expansion
results.

In Table~\ref{tab:recfns} we present our results of the all-order (in $Z\alpha$) calculation
in comparison with
the sum of the $\Za$-expansion contributions up to $(\Za)^6$. We observe
excellent agreement of the two methods in the low-$Z$ region. By contrast, for high $Z$ the all-order
results become larger than the $\Za$-expansion estimates by an order of magnitude.
In the last column of Table~\ref{tab:recfns} results of previous approximate treatment
\cite{shabaev:98:recground,yerokhin:15:recprl,yerokhin:16:recoil} are listed (recalculated for the nuclear model
and nuclear radii adopted in this work). The previous treatment was incomplete
because the transverse part of the finite-size photon propagator was not known at that
time. As seen from the table, this incompleteness leads to effects ranging from 1.5\% for $Z = 1$ to 9\% for
$Z = 92$.

\begin{table}
\caption{Finite-size nuclear recoil correction for the $1s$ state of H-like ions,
to be multiplied by the prefactor $(m^2/M) (\Za)^5/\pi$.
\label{tab:recfns}}
\begin{ruledtabular}
\begin{tabular}{lcw{2.6}w{2.6}w{2.6}}
\multicolumn{1}{l}{$Z$}
& \multicolumn{1}{l}{$r_C$ [fm]}
        & \multicolumn{1}{c}{$\Za$-expansion}
        & \multicolumn{1}{c}{All-order}
        & \multicolumn{1}{c}{Refs. \cite{shabaev:98:recground,yerokhin:16:recoil}}
  \\
\hline\\[-5pt]
  1 & 0.8409 &  -0.00419 & -0.00419 & -0.00425 \\
  2 & 1.6755 &  -0.00849 & -0.00850 & -0.00874\\
  3 & 2.4440 &  -0.01229 & -0.01233 & -0.01281\\
  5 & 2.4060 &  -0.00775 & -0.00782 & -0.00829\\
 10 & 3.0055 &  -0.00752 & -0.00780 & -0.00850\\
 20 & 3.4776 &  -0.00829 & -0.00962 & -0.01042\\
 30 & 3.9283 &  -0.01079 & -0.01506 & -0.01572\\
 40 & 4.2694 &  -0.0137  & -0.0246  & -0.0247\\
 50 & 4.6519 &  -0.0174  & -0.0429  & -0.0414 \\
 60 & 4.9123 &  -0.0210  & -0.0764  & -0.0717 \\
 70 & 5.3108 &  -0.0258  & -0.148   & -0.137\\
 80 & 5.4648 &  -0.0296  & -0.298   & -0.274\\
 92 & 5.8571 &  -0.0358  & -0.819   & -0.757\\
\end{tabular}
\end{ruledtabular}
\end{table}

{\sl Conclusions.---}
In this Letter we performed rigorous QED calculations of the finite-size nuclear recoil (recfns) effect
for the Lamb shift of hydrogen-like atoms, both within the $\Za$-expansion and
to all orders in $\Za$. The resulting correction for the $1S$-$2S$ transition frequency in hydrogen
is $-1.62$~kHz, which may be compared with the experimental uncertainty of
$0.01$~kHz \cite{parthey:11,matveev:13} in hydrogen, $5.4$~kHz \cite{ahmadi:18} in antihydrogen,
and the total theoretical uncertainty of $1.6$~kHz \cite{tiesinga:21:codata18}. The higher-order $(\Za)^{5+}$ contribution is quite small
for light ions ($-0.04$~kHz for the $1S$-$2S$ transition in hydrogen) but becomes increasingly important with
growth of $Z$. Generally, the recfns correction is comparable in magnitude with the nuclear-structure effects and should be
included into consideration for obtaining high-precision theoretical predictions of the Lamb shift.
In particular, the recfns effect contributes to nonlinearities of the so-called King's plots, which are
nowadays considered as a promising tool for searches for new particles \cite{berengut:18,flambaum:18}.

The developed approach for describing the recoil effect with a finite-size nucleus
to all orders in $\Za$ may find many applications in precision studies of simple atomic systems.
It will lead to more accurate theoretical predictions of the bound-electron $g$-factor
and to improved spectra of muonic atoms. In particular, it opens a way to
a non-perturbative treatment of the vacuum-polarization combined with the nuclear recoil in muonic
atoms. More specifically, in muonic atoms the vacuum-polarization, the nuclear recoil,
and the fns effects are of comparable magnitude and are difficult to be
accounted for by perturbation theory. Our approach allows one to account for the nuclear recoil
modified not only by the fns but also by the Uehling vacuum-polarization,
without any expansion in $\Za$, which has not been accomplished so far
\cite{diepold:18,franke:17}. Furthermore, the developed approach can be used for
deriving the exact (in $\Za$) formulas for the recoil effect to the hyperfine splitting,
which is presently unknown for medium- and high-$Z$ electronic and muonic atoms.

K.P. acknowledges support from the National Science Center (Poland) Grant No. 2017/27/B/ST2/02459.
V.A.Y. was supported by the Russian Science Foundation Grant No. 20-62-46006.


\end{document}